\documentclass[%
 reprint,
 amsmath,amssymb,
 aps,
]{revtex4-1}

\usepackage{graphicx}
\usepackage{dcolumn}
\usepackage{bm}

\begin{document}

\preprint{APS/123-QED}

\title{The Real Quaternion Relativity}
\thanks{Thanks to:  Prof. Leon Altschul for
	helpful discussions}%
\author{Viktor Ariel}
\affiliation{%
}%

\begin{abstract}
In this work, we use real quaternions and the basic concept of the final speed of light in an attempt to enhance the standard description of special relativity.  First, we demonstrate that it is possible to introduce a quaternion time domain where a coordinate point is described by a quaternion time. We show that the time measurement is a function of the observer location, even for stationary frames of reference. We introduce a moving observer, which leads to the traditional Lorentz relation for the time interval. We show that the present approach can be used in stationary, moving, or rotating frames of reference, unlike the traditional special relativity, which applies only to the inertial moving frames. Then, we use the quaternion formulation of space-time and mass-energy equivalence to extend the quaternion relativity to space, mass, and energy. We demonstrate that the transition between the particle and observer reference frames is equivalent to space inversion and can be described mathematically by quaternion conjugation. On the other hand,  physical measurements are described by the quaternion norm and consequently are independent of the quaternion conjugation, which seems to be the quaternion formulation of the relativity principle.  \\
\end{abstract}

\pacs{Valid PACS appear here}
\maketitle


\section{\label{sec:level1}Introduction\protect\\}  
 
William Rowan Hamilton, who invented quaternions \cite{Hamilton}, once famously observed "how the One of Time, of Space the Three,
might in the chain of symbols girdled be" \cite{Graves}. It appears that he anticipated  the modern four-dimensional space-time based on his quaternion discovery. Unfortunately, the quaternion approach was not used by Einstein while developing the special theory of relativity \cite{Einstein} despite apparent advantages of using quaternions over Minkowski's four-vectors \cite{Ariel}. Bi-quaternions were applied to special relativity \cite{Silberstein} and showed initial promise in developing a unified field theory \cite{Gsponer}, however bi-quaternions do not form a division algebra and consequently lead to mathematical difficulties. It was shown that it is possible to use the finite speed of light propagation and the Doppler effect for an alternative mathematical derivation of the relativity equations \cite{Moriconi}. Previously, we used the real quaternions for the time and mass by a heuristic comparison between the Minkowski's and quaternion mathematical formulations \cite{Ariel}.

In this work, we use the basic principle of the finite speed of light propagation and the algebra of real quaternions to derive the principle of relativity. First, we define the quaternion time domain and a stationary-clock reference system where a clock is located at the zero-point of a three-dimensional time vector. We introduce a stationary observer, which can record the clock time signals using a watch resulting in scalar time interval measurements.

We demonstrate that there is a time delay between the signal sent by the clock and the signal recorded by the observer due to the final speed of light propagation from the clock to the observer. While this result is expected, we show that the real quaternions form a beautiful mathematical formalism for the description of the underlying phenomena. Applying the above approach to a moving observer together with a quaternion definition of the velocity results in the relativistic Lorentz relation for time. We show that the experimental time measurements are expressed mathematically by the quaternion norm and consequently are independent of the quaternion conjugation, which seems to be the mathematical statement of the principle of relativity. We show that the same quaternion approach can be applied to the treatment of particle mass. Finally, we extend the quaternion relativity description to space and energy using the quaternion space-time and mass-energy equivalence.

\section{Quaternion Time and Space-Time Equivalence} %

\begin{figure}
	\includegraphics{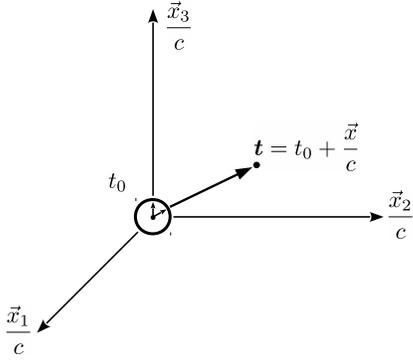}
	\caption{\label{fig:clock_frame} Define the quaternion time domain with the clock located at the zero-point.}
\end{figure}

Let us consider a quaternion time domain with a scalar clock, $t_0$, located at the zero-point of a three-dimensional coordinate system, $(\vec{x}_1/c,\thinspace \vec{x}_2/c,\thinspace  \vec{x}_3/c)$, as shown in Fig.~\ref{fig:clock_frame}. Here, $c$ is the speed of light in vacuum, which we consider a scalar constant. The quaternion notation used in this work is presented in the appendix. 

Since the clock is stationary and located at the zero-point of the coordinate system, we call this system - the clock reference frame. Then, an arbitrary point located in our quaternion time domain is described mathematically by the real quaternion coordinates, 

\begin{equation}
		\boldsymbol{ t} 
		=  t_0+\thinspace \dfrac {\vec{  x} } {c} \thinspace .		
	\label{eq:q_time}
\end{equation}
Note that once we defined the clock reference system, we introduced an asymmetry into the description of the three-dimensional space domain, since now every point in space will receive the clock signal in the future due to the finite speed of light propagation. Also, note that $\vec{x}/c$ in (\ref{eq:q_time}) describes the time of light signal propagation from the clock to the space coordinate $\vec x$, thus only at the zero-point one can potentially measure the accurate time of the clock.

\begin{figure}
	\includegraphics{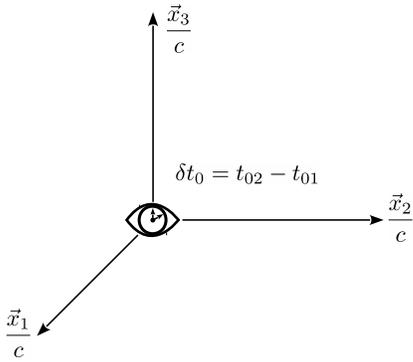}
	\caption{\label{fig:observer_frame}Adding an observer at the zero-point of the time domain.}
\end{figure}

Let us place an observer at the same location as the clock, which is the zero-point of the clock reference system, as can be seen in Fig.~\ref{fig:observer_frame}. Since the distance between the observer and the clock is zero, the observer can measure the scalar time of the clock instantaneously resulting in the accurate scalar time measurement of the clock, $\boldsymbol t = t_0$. In general, any observer at the zero-point measures scalar time quantities only.

For example if $t_{01}$ and $t_{02}$ are two clock signals at the zero-point, the observer can measure the scalar time interval, 
\begin{equation}
\delta t_0 
=  t_{02} - t_{01} \thinspace .		
\label{eq:q_time_interval}
\end{equation}

\begin{figure}
	\includegraphics{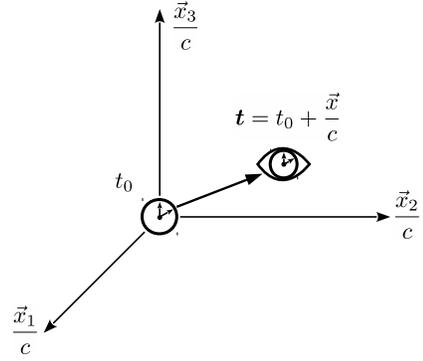}
	\caption{\label{fig:q_clock_observer} A stationary observer in the clock reference frame.}
\end{figure}

Let us now place a stationary observer at a quaternion location $\boldsymbol t$, while the clock remains at the zero-point, as in Fig.~\ref{fig:q_clock_observer}. Let us assume that the observer has a watch located at the exactly the same location as the observer and which is perfectly synchronized with the zero-point clock.

We choose the direction of vector $\vec{x}$ according to the direction of light propagation from the clock to the observer. Thus, the observer looking at the zero-point clock and simultaneously checking the watch will see two different times due to the delay in light propagation from the clock to the observer.

Thus, the observer has two simultaneous time measurements:  a scalar time image received from the zero-point clock, $t_0$, and a scalar time measured on the observer watch, $t$, recorded when the signal $t_0$ arrives at the observer location. Therefore, the observer can calculate the time interval of the light propagation from the clock to the observer. 

\begin{figure}
	\includegraphics{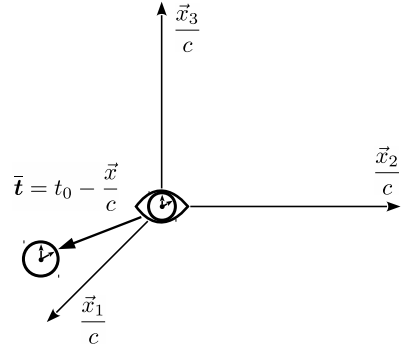}
	\caption{\label{fig:q_observer_clock} In the observer reference frame, the location of the clock is inverted in comparison with the clock reference frame.}
\end{figure}

Assuming that the time quaternion at the observer is described by
(\ref{eq:q_time}), while the quaternion at the clock is a scalar given by $\boldsymbol {t} = t_0$, we calculate the quaternion time interval,
\begin{equation}
\boldsymbol{ \delta t} 
=\boldsymbol{ t}  - \boldsymbol{t_0} 
=  t_0+\thinspace \dfrac {\vec{  x} } {c} -  t_0 = \dfrac {\vec{  x} } {c}\thinspace ,		
\label{eq:q_time_difference}
\end{equation}
which is the time vector of light propagation.
Clearly, the measured time interval should be a scalar independent of the direction of light propagation and only a function of the distance between the clock and the observer. Therefore, we assume that the measured time interval is described by the scalar norm of the quaternion (\ref{eq:q_time_difference}),  
\begin{equation}
 \delta t
=  t - t_0
=\sqrt {\boldsymbol {\delta t} \thinspace 
	\bar {\boldsymbol {\delta t}}} 
= \sqrt {\thinspace  \dfrac {{  x \thinspace}^2} {c^2}}
=\dfrac {{  x} } {c}\thinspace .		
\label{eq:q_time_diff_measure}
\end{equation}

Thus, we obtained a scalar time difference between the clock and the watch due to the finite time of light propagation between the clock and the observer. 

For example, consider a clock hanging on a kitchen wall while the observer is 3m away. The measured time difference between the kitchen clock and the observer's watch will be 10ns, which can be easily recorded by a modern measuring instrument. 

We can introduce several synchronized clocks at various distances from the observer resulting in different time interval measurements by the observer. Therefore, it would be convenient to introduce the observer reference frame.

Let us then repeat the previous experiment in the observer frame, where the observer with a watch is located at the zero-point, while the vector coordinate system is the same as in the clock frame, as seen in Fig.~\ref{fig:q_observer_clock}. 

Now, the observer looks at the watch and sees the time, $t_0$, while simultaneously looking at the clock and recording the time, $t < t_0$. Note that the location of the clock is described by the vector, $-\vec{x}/c$, and the time on the clock is described by a conjugate time quaternion,
\begin{equation}
\boldsymbol{\bar{t}}
=   t_0-\thinspace \dfrac {\vec {  x }} {c} \thinspace.		
\label{eq:conj_q_time}
\end{equation}
Note that the scalar time on the observer watch, $t_0$, corresponds to the zero-point location of the observer in the observer frame. 

This leads to a conjugate quaternion time difference, 
\begin{equation}
\bar{\boldsymbol{ \delta t}} 
=  t_0-\thinspace \dfrac {\vec{  x} } {c} -  t_0 = -\dfrac {\vec{  x} } {c}\thinspace .		
\label{eq:q_conj_time_diff}
\end{equation}

Again, we assume that the measured time interval is described by the quaternion norm of (\ref{eq:q_conj_time_diff}) giving the same experimental measurement as in the clock reference frame,
\begin{equation}
\delta t = \sqrt {\boldsymbol{ \delta t}\thinspace \boldsymbol{\bar{ \delta t}}}
= \sqrt {\boldsymbol{\bar{ \delta t}} \boldsymbol{ \delta t}\thinspace }
= \sqrt {\thinspace  \dfrac {{  x \thinspace}^2} {c^2}} = \dfrac {  x} {c}\thinspace .
\label{eq:scalar_time_diff}
\end{equation} 
As expected, we measured the same time difference in both stationary frames of reference and expressed it by  (\ref{eq:q_time_diff_measure}) and (\ref{eq:scalar_time_diff}). Also note that we can use the time interval in (\ref{eq:q_time_diff_measure}) and (\ref{eq:scalar_time_diff}) to calculate the distance between the clock and observer, $x = c \thinspace \delta t$.

It appears from (\ref{eq:conj_q_time}) and (\ref{eq:q_conj_time_diff}) that the observer reference frame introduces an asymmetry in the time domain, since the observer always receives the time signals from the past. To emphasize, an observer in the observer reference frame is always the last one to observe all the available clock signals from the universe. Also in our quaternion formulation, the clock and observer reference frames are not mathematically equivalent since we need to perform space inversion in transition from one to another. If we use regular quaternions in the clock reference frame, we need to use conjugate quaternions for the time intervals in the observer reference frame. Therefore, by introducing a clock and an observer we created a fundamental asymmetry in the time domain due to the finite speed of signal propagation. It seems possible to say that a clock-observer pair creates asymmetric quaternion arrows of time as in Fig.~\ref{fig:q_clock_observer} and Fig.~\ref{fig:q_observer_clock}. 

We assume that there are no additional experimental uncertainties and measurement delays, which results in the equal sign in (\ref{eq:scalar_time_diff}) and everywhere else in this work. This is clearly a very optimistic assumption and in practice we should expect experimental time delays. Therefore, we assume that all the expressions in this work are the optimal values obtained under the ideal measurement conditions.

\begin{figure}
	\includegraphics{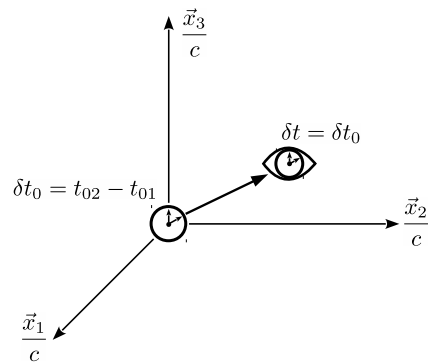}
	\caption{\label{fig:observer_time_interval}Time interval at the stationary observer location is the same as at the zero-point.}
\end{figure}

Let us now consider the measurement of a time interval between two different time signals from the zero-point clock, $\delta t_0 = t_{02} - t_{01}$, as seen in Fig.~\ref{fig:observer_time_interval}. In the clock reference frame, the time signals at the observer location are described by quaternions,
\begin{equation}
\begin{cases}
\boldsymbol{t_1} = t_{01} + \dfrac {\vec{x}}  {c}  
\\
\\
\boldsymbol{t_2} = t_{02} + \dfrac {\vec{x}}  {c}  
\end{cases}
\label{eq:q_time_interval_def}
\end{equation}
Note that for a stationary observer the time vectors of the two quaternions in (\ref{eq:q_time_interval_def})
are the same. The measurement of the quaternion time interval by the observer using (\ref{eq:q_time_interval_def}) will result in,
\begin{equation}
{\boldsymbol{ \delta t}} 
=  t_{02} - t_{01} + \thinspace \dfrac {\vec{  x} } {c} - \thinspace \dfrac {\vec{  x} } {c}   =  t_{02} - t_{01} = \delta t\thinspace.		
\label{eq:q_time_interval_observer}
\end{equation}
Therefore, the time interval measured by the observer is the same as the time interval at the clock due to the cancellation of the light propagation delay for a stationary observer.

\begin{figure}
	\includegraphics{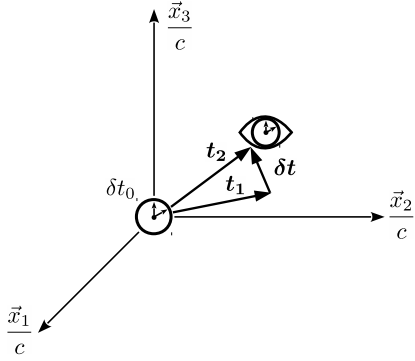}
	\caption{\label{fig:clock_velocity}The observer is moving relative to the stationary clock located at the zero-point.}
\end{figure}

Let us now consider a moving observer, which during the clock time interval, $\delta t_0 = t_{02} - t_{01}$, moved by a space interval, $\delta \vec{x} = x_2 - x_1 $, as shown in Fig.~\ref{fig:clock_velocity}.

We can define the initial and final quaternion time values as in (\ref{eq:q_time_interval_def}), however, the initial and final vector times are no longer the same,

\begin{equation}
\begin{cases}
\boldsymbol{t_1} = t_{01} + \dfrac {\vec{x_1}}  {c}  
\\
\\
\boldsymbol{t_2} = t_{02} + \dfrac {\vec{x_2}}  {c}  
\end{cases}
\label{eq:q_time_interval_move} 
\end{equation}

We can now define the quaternion time interval at the moving observer using (\ref{eq:q_time_interval_move}),
\begin{equation}
{\boldsymbol{ \delta t}} 
=  \boldsymbol{t_{2}} - \boldsymbol{t_{1}}
=  t_{02} - t_{01} + \thinspace \dfrac {\vec{  x_2} } {c} - \thinspace \dfrac {\vec{  x_1} } {c}  
= \delta t_0 + \thinspace \dfrac {\vec{  \delta x} } {c} \thinspace.		
\label{eq:q_time_interval_observer_move}
\end{equation}
Let us define the average quaternion velocity of the moving observer \cite{Ariel} during the time interval $\boldsymbol {\delta t}$ as,
\begin{equation}
\boldsymbol {v} = \dfrac {\delta \vec{x}}  {\boldsymbol {\delta t}} \thinspace.		
\label{eq:q_velocity_define}
\end{equation}
Note that in (\ref{eq:q_velocity_define}) we used the fact that the quaternion division is well defined in the real quaternion algebra.

We can now express the observer time interval $\boldsymbol{\delta t}$ in the clock reference frame using (\ref{eq:q_time_interval_observer_move}) and (\ref{eq:q_velocity_define}), 

\begin{equation}
	\boldsymbol{\delta t} 
	=  \delta t_0+\thinspace \dfrac {\vec {\delta  x }} {c}
	=  \delta t_0+\thinspace \dfrac {\boldsymbol{v} \thinspace 
		\boldsymbol{\delta t}} {c} \thinspace .
	\label{eq:quat_time_interval}
\end{equation}
Thus, we obtained a recursive quaternion relation that describes the time interval in terms of its zero-point value together with the vector displacement, which is expressed in terms of the same time interval and velocity. We can consider the vector displacement as a correction to the measured time interval which is a function of the normalized velocity $\boldsymbol {\beta} = \boldsymbol {v} / c$. Note that we did not make any assumptions about the direction of the velocity, therefore, the quaternion interval in  (\ref{eq:quat_time_interval}) seems to describe both  the translational and rotational types of motion.

\begin{figure}
	\includegraphics{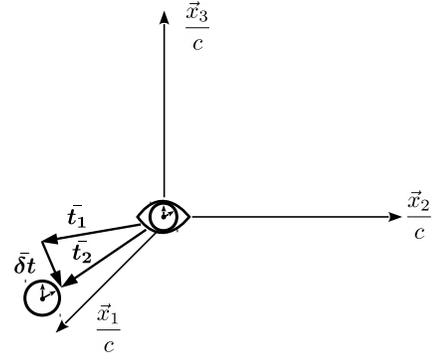}
	\caption{\label{fig:observer_velocity}In the observer reference frame, the movement of the clock is inverted in comparison with the clock reference frame and is described by conjugate quaternions.}
\end{figure}

We can now repeat the experiment in the observer reference frame where the clock is moving, as shown in Fig.~\ref{fig:observer_velocity}. As can be seen in the figure, the clock is moving in the opposite direction from the observer. This is certainly expected since a stationary clock on a train platform, as seen by an observer located on the train, moves in the opposite direction to the movement of the train. Therefore, we can describe the quaternion time interval at the moving clock similar to (\ref{eq:quat_time_interval}) but using a conjugate quaternion time interval,

\begin{equation}
	\boldsymbol{\bar{\delta t}}
	=  \delta t_0-\thinspace \dfrac {\vec {\delta  x } } {c}
	=  \delta t_0-\thinspace \dfrac {\boldsymbol{v} \thinspace
		\boldsymbol{\delta t}} {c} \thinspace .
	\label{eq:conj_time_interval}
\end{equation}

As usual, the result of the experiment should not depend on the reference frame and should provide the same value of the time interval in the clock and observer reference frames. Furthermore, the measurement should not depend on the vector direction but only on its absolute value. Therefore, we expect that the measured time interval depends on the quaternion norm only,  

\begin{equation}
\delta t^2 = {\boldsymbol{ \delta t}\thinspace \boldsymbol{\bar{ \delta t}}}
= {\delta t_0^2
	+\thinspace  \dfrac {{  \delta x \thinspace}^2} {c^2}} 
= {\delta t_0^2
	+\thinspace  \dfrac {  v^2 \delta t^2  \thinspace} {c^2}} \thinspace .
\label{eq:q_interval_norm}
\end{equation} 

This leads to the traditional form of the Lorentz relation for the moving time interval in both the clock and observer reference frames,
\begin{equation}
\delta t
= \dfrac {\delta t_0}
{\sqrt {1- \dfrac { v^2} {c^2} }}\thinspace.
\label{eq:t_lorentz}
\end{equation}   
While the Lorentz relation (\ref{eq:t_lorentz}) is similar to the result of the traditional special relativity, we showed here that the time dilation in (\ref{eq:t_lorentz}) is fundamentally related to the basic concept of the finite speed of light and can be derived using the algebra of real quaternions. While it is possible to claim, based on (\ref{eq:t_lorentz}), that the time is different in different moving frames of reference, it seems more physically accurate to say the time measurement depends on the relative position and movement between the observer and the clock. 

To summarize, we expressed the time by real quaternions in the clock reference frame and by conjugate quaternions in the observer reference frame. The measured time and time intervals depend on the location and relative movement between the clock and observer but independent of the frame of reference. We use quaternions to describe time and time intervals in the clock reference frame and we use conjugate quaternions in the observer frame. The measured time intervals are described by the quaternion norm and consequently are the same in the clock and in the observer reference frames.  

We can formulate the quaternion principle of relativity by saying that quaternion conjugation describes the transition between the clock and  observer reference frames, while the quaternion norm describes the measurements, which implies that the measurements are independent of the reference frames. 

We did not us the assumption of the inertial frames of reference, unlike in Einstein's relativity, therefore our quaternion approach seems suitable for the description of relativity in arbitrary frames of reference: stationary, moving, or rotating.

\section{Quaternion Space-Time and Mass-Energy Equivalence} %

Our quaternion time domain can be visualized as a map describing how long it takes for the time signal to propagate from the clock zero-point to the current space location.  We have a clear mechanism for distance interval measurements $\delta x = c\thinspace \delta t$ using clocks and observers from (\ref{eq:q_time_diff_measure}), (\ref{eq:scalar_time_diff}),  and (\ref{eq:t_lorentz}). 

Therefore, we can multiply our time quaternions by the speed of light and obtain the definition of the quaternion space in both the clock and observer reference frames,
\begin{equation}
\begin{cases}
\boldsymbol{ s} 
=c\thinspace \boldsymbol{ t}
=  c \thinspace t_0+\thinspace {\vec{  x} }
\\
\\
\boldsymbol{\bar { s}} 
=c\thinspace \boldsymbol{ t}
=  c \thinspace t_0-\thinspace {\vec{  x} }
\end{cases}
\label{eq:q_space-time}
\end{equation}  
Thus, (\ref{eq:q_space-time}) can be viewed as a mathematical expression of the quaternion space-time equivalence due to the fact that distances can always be measured with clocks and observers. The most common practical example of this equivalence it the modern global positioning system.

Let us consider a free particle with a zero-point mass, $m_0$, and a vector momentum $\vec {p} $. Then, we can introduce a quaternion mass domain \cite{Ariel} using the same approach as we applied to time intervals, namely we modify the zero-point mass by a relativistic mass correction, which depends on the ratio of the velocity to the speed of light,
\begin{equation}
\begin{cases}
\boldsymbol{m} =   m_0
+\thinspace \dfrac {\vec p } {c}
=  m_0+\thinspace \dfrac {\boldsymbol{m v}} {c}
\\
\\
\boldsymbol{ \bar{m}} =   m_0
-\thinspace \dfrac { \vec p } {c}
=  m_0-\thinspace \dfrac {\boldsymbol{ m v}} {c} 
\end{cases}
\label{eq:q_mass}
\end{equation}  

\begin{figure}
	\includegraphics{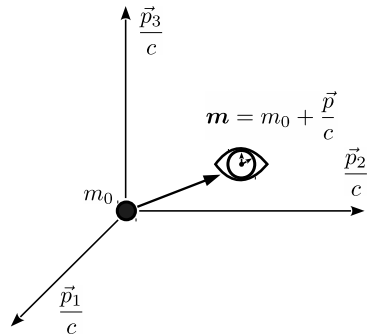}
	\caption{\label{fig:q_mass_observer} Quaternion representation of the mass of a body in the body reference frame.}
\end{figure}

\begin{figure}
	\includegraphics{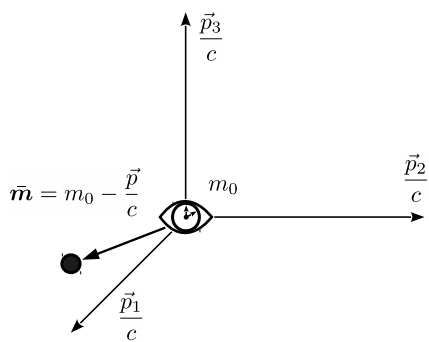}
	\caption{\label{fig:q_observer_mass} Quaternion representation of mass in the observer reference frame.}
\end{figure}

The result is represented  in the particles reference frame together with a moving observer in Fig.~\ref{fig:q_mass_observer}, while the particle is shown in the observer frame in Fig.~\ref{fig:q_observer_mass}. 

In (\ref{eq:q_mass}), we used the quaternion definition of the particle momentum, $\vec {p} = {\boldsymbol{m v}}$ \cite{Ariel}. Similar to the quaternion time, we use a regular quaternion for the mass description in the particle frame, while we use a conjugate mass quaternion in the observer reference frame. This seems justified because of the space inversion and the change in the quaternion velocity direction in (\ref{eq:q_mass}) between the particle and the observer frames.

Note that the particle mass measured in the observer frame is equal to the zero-point mass, $m_0$,  while it is equal to $m$ for a moving observer. 
We assume that, $m$,  is expressed by the norm of the quaternion mass and is the same in the particle and observer frames and can be calculated from,
\begin{equation}
m^2 = \boldsymbol{m}\thinspace \boldsymbol{\bar{m}}
= m_0^2
+\thinspace  \dfrac {  p^2} {c^2} 
= m_0^2
+\thinspace  \dfrac {v^2} {c^2} m^2\thinspace.
\label{eq:m_norm_squared}
\end{equation}  

The Lorentz relation for the mass of a moving particle is easily obtained from (\ref {eq:m_norm_squared}), 

\begin{equation}
m
= \dfrac {m_0}
{\sqrt {1-\dfrac { v^2} {c^2} }}\thinspace.
\label{eq:m_lorentz_norm}
\end{equation}   
Therefore, the Lorentz relation in the mass domain applies to the absolute value of the particle mass similar to time intervals in the time domain (\ref{eq:t_lorentz}).

Let us now consider a quaternion mass-energy equivalence and define the total energy, $\boldsymbol{\epsilon}$, of a free  particle,
\begin{equation}
\begin{cases}
\boldsymbol{\epsilon}= \boldsymbol{m}c^2 =   m_0c^2
+\thinspace {\vec p } {c}

\\
\\
\boldsymbol{\bar{\epsilon}}=  \boldsymbol{ \bar{m}}c^2 =   m_0c^2
-\thinspace  { \vec p } {c}

\end{cases}
\label{eq:q_energy-mass}
\end{equation}

Here, we can define the zero-point particle energy as $\epsilon_0 = m_0c^2$.
The measured scalar value of the quaternion energy can be calculated from the squared norm,

\begin{equation}
\epsilon^2 
= m_0^2 c^4
+\thinspace  {p\thinspace^2} {c^2}  \thinspace ,
\label{eq:e_measured}
\end{equation}  
which  we recognize as the relativistic energy-momentum dispersion relation for a free particle.  Assuming relatively large zero-point energy, we can approximate the difference between the moving free particle energy and its zero-point value,
\begin{equation}
\delta\epsilon = \epsilon - \epsilon_0
=\sqrt{ \epsilon_0^2
	+\thinspace  {p\thinspace^2} {c^2} } - \epsilon_0
\simeq \dfrac {p^2} {2m_0} \thinspace ,
\label{eq:m_lorentz}
\end{equation}  
which is the classical kinetic energy of the particle.

Thus, we were able to define the quaternion mass and mass-energy equivalence by using real quaternions and  the basic principle of the finite speed of light.

\section{Conclusions}
In this work, we used the basic concept of the finite speed of light propagation and the real quaternion mathematical formalism to derive an enhanced formulation of the special relativity. We show that this approach leads to different time measurements for different locations of the clock and observer even in stationary frames of reference. By using a moving observer, we derived the traditional Lorentz relations for the time intervals and free particle mass.  We demonstrated that the present approach can be used in stationary, moving, or rotating frames of reference unlike the traditional special relativity, which only works for the inertial frames of reference. We extended the quaternion relativistic approach to space, mass, and energy using the concepts of quaternion space-time and mass-energy equivalence. We demonstrated that if we use regular quaternions in the particle frame of reference,  we need to use conjugate quaternions in the observer frame. Therefore, space inversion is used in transition between the frames of reference, which is mathematically equivalent to quaternion conjugation. On the other hand, the measured results are based on the quaternion norm and consequently are independent of the quaternion conjugation. This appears to be the mathematical statement of the quaternion relativity principle.

\section{Appendix}
In this work, a real quaternion, $\textit{\textbf a}$, and its conjugate, $\bar{\textit{\textbf a}}$, are written in terms of a scalar, $a_0$, and an imaginary vector, $\vec{a}$,

\begin{equation}
\begin{cases}
{\textit{\textbf a}}  =a_0+ \thinspace\vec{a} = 
a_0+ \vec{i_1}\thinspace a_1 +  \vec{i_2}\thinspace a_2 + \vec{i_3}\thinspace a_3 \thinspace 
\\
\bar{{\textit{\textbf a}}}  =a_0- \thinspace\vec{a} = 
a_0- \vec{i_1}\thinspace a_1 -  \vec{i_2}\thinspace a_2 - \vec{i_3}\thinspace a_3 \thinspace 
\end{cases}
\label{eq:define_quaternion}
\end{equation}
The Euclidian imaginary vector basis is defined according to Hamilton \cite{Hamilton} as $\vec{i}\thinspace_1^2=\vec{i}\thinspace_2^2=\vec{i}\thinspace_3^2= \vec{i}_1\thinspace \vec{i}_2\thinspace \vec{i}_3 = -1$. The Euclidean norm of a quaternion is a scalar quantity defined in terms of the quaternion inner product as,

\begin{equation}
a = \|{\textit{\textbf a}}\|
=\sqrt{ \textit{\textbf a}\cdot \bar{\textit{\textbf a}}}\thinspace
=\sqrt{ \bar{\textit{\textbf a}}\cdot {\textit{\textbf a}}}\thinspace . 
\label{eq:define_abs}
\end{equation}

In general, we write quaternions using bold characters and quaternion absolute values and scalars as regular characters. We use the vector sign above vectors similar to Gibbs-Heaviside notation, however, all vectors in this work are considered imaginary as originally intended by Hamilton.

\end{document}